\documentclass[%
 reprint,
 showkeys,
superscriptaddress,
longbibliography,
 amsmath,amssymb,
 aps,
floatfix
]{revtex4-1}

\usepackage{tabularx,booktabs}
\usepackage{graphicx}
\usepackage{dcolumn}

\usepackage{bm}
\usepackage[colorlinks = true,
            linkcolor = blue,
            urlcolor  = blue,
            citecolor = blue,
            anchorcolor = blue]{hyperref}
\usepackage{siunitx}

\usepackage{xcolor}

\begin{document}

\title{One-Dimensional Disordered Bosonic Systems}

\author{Chiara D'Errico}\email{chiara.derrico@cnr.it}
\affiliation{Istituto per la Protezione Sostenibile delle Piante, CNR-IPSP, 10135 Torino, Italy}
\affiliation{European Laboratory for Non-Linear Spectroscopy, LENS, 50019 Sesto Fiorentino, Italy}
\author{Marco G. Tarallo}
\affiliation{Istituto Nazionale di Ricerca Metrologica, Strada delle Cacce 91, 10135 Torino, Italy}

\begin{abstract}
Disorder is everywhere in nature and it has a fundamental impact on the behavior of many quantum systems. The presence of a small amount of disorder, in fact, can dramatically change the coherence and transport properties of a system. Despite the growing interest in this topic, a complete understanding of the issue is still missing. 
An open question, for example, is the description of the interplay of disorder and interactions, which has been predicted to give rise to exotic states of matter such as quantum glasses or many-body localization. In this review, we will present an overview of experimental observations with disordered quantum gases, focused on one-dimensional bosons, and we will connect them with theoretical predictions.
\end{abstract}


\maketitle

\section{Introduction}
Ultracold atoms platforms are able to mimic the physics of other quantum many-body systems \cite{Bloch2012,Gross2017,Schafer2020}. Thanks to the high degree of tunability of many important parameters, they have been used to study the low-temperature quantum phases and the transport properties of neutral particles with short-range interaction \cite{Lewenstein2007,Bloch2008}.
Their strong versatility allows researchers to use these platforms to investigate the physics of disorder \cite{Sanchez-Palencia2010,Modugno_2010,Shapiro2012}, mainly using two different kinds of optical disordered potentials: laser 
speckles \cite{Lye2005,Clement2005,Fort2005,Schulte2005,Clement2006,Chen2008,Billy2008,White2009,Pasienski2010,Dries2010,Josse2018,Josse2019} and quasiperiodic lattices \cite{Fallani2007,Guarrera2007,Guarrera2008,Roati2008,Deissler2010,Deissler2011,Lucioni2011,DErrico2013,Lucioni2013,DErrico_AIP,Gadway2011,Tanzi2013,DErrico2014,Gori2016}, both allowing for the first observation of Anderson localization in matter-waves \cite{Billy2008,Roati2008}. Although the present review is devoted to one-dimensional (1D) bosons, it is important to mention that the possibility to control the dimensionality of the systems allowed experimentalists to also study 2D diffusion \cite{Bouyer2010} and coherence \cite{Allard2012}, coherent backscattering \cite{Josse2012,Josse2015}, and 3D Anderson localization with both fermions \cite{Kondov2011} and bosons \cite{Jendrzejewski2012,Semeghini2015}.

Despite many years of investigation and the many efforts that have been undertaken, both from the experimental and theoretical point of view, a clear and complete characterization of the effect of disorder on transport and coherence of a quantum system is still missing.
An open issue, for example, is the description of the non-trivial interplay between disorder and interactions, which has been predicted to give rise to exotic states of matter such as quantum glasses \cite{Giamarchi&Schulz,Fisher&Fisher} or many-body localization \cite{Aleiner2010,Iyer2013}. In particular, a transition between a superfluid phase for weakly repulsive bosons and a localized Bose glass phase for strong repulsion has been predicted both for one-dimensional \cite{Giamarchi&Schulz} and higher-dimensional \cite{Fisher&Fisher} bosons. However, the first experimental attempts to insert weak interactions in Anderson-localized disordered systems have clearly shown that the interaction energy can compete with disorder and induce delocalization by restoring coherence \cite{Deissler2010,Deissler2011} or transport \mbox{\cite{Lucioni2011,DErrico2013,Lucioni2013,DErrico_AIP}}. The quest for the effect of strong interactions requires to freeze the radial degrees of freedom, for example, by reducing the dimensionality of the system. One-dimensional bosons are the prototype disordered systems, with an established theoretical framework, useful to answer to some of the fundamental questions about the quantum phases and the transport properties of low-temperature matter.

In this review, we will focus on the experimental observations obtained with ultracold quantum gases \cite{Fallani2007,Guarrera2007,Guarrera2008,Gadway2011,Tanzi2013,DErrico2014,Gori2016}. In particular, after a brief survey of the theoretical background of 1D disordered systems, we review the experimental results achieved to detect and study disordered interacting quantum phases, analyzing their signature on coherence, transport, and energy excitation properties.

\section{Theoretical Background of 1D Disordered Systems}\label{sec:Theo}

Let us consider a disordered Bose gas in a discrete 1D space, whose space dependence is described by the site index $j$.
This system is described by a modified Bose--Hubbard Hamiltonian:
\begin{eqnarray}\label{eq:BHdis}
    H &=& -J\sum_j (b^\dagger_{j+1}b_j +b^\dagger_jb_{j+1})+\frac{U}{2}\sum_jn_j(n_j-1)\nonumber\\
    &&+\sum_j (\epsilon_j+V^{\rm{HO}}_j)n_j
\end{eqnarray}
where $b_j$ denotes the boson annihilation operator at site $j$, while the site occupation quantified by the usual operator $n_j =b_j^\dagger b_j$. The first two terms on the right-hand side of Equation~\eqref{eq:BHdis} represent the usual Bose--Hubbard interactions, corresponding to site-to-site tunneling with a rate $J$ and the on-site repulsion $(U>0)$. The third term in Equation~\eqref{eq:BHdis} accounts for the presence of both the harmonic trap $V^{\rm{HO}}_j$ and the disorder potential $\epsilon_j$.

\begin{figure}[t]
    \centering
    \includegraphics[width=0.4\textwidth]{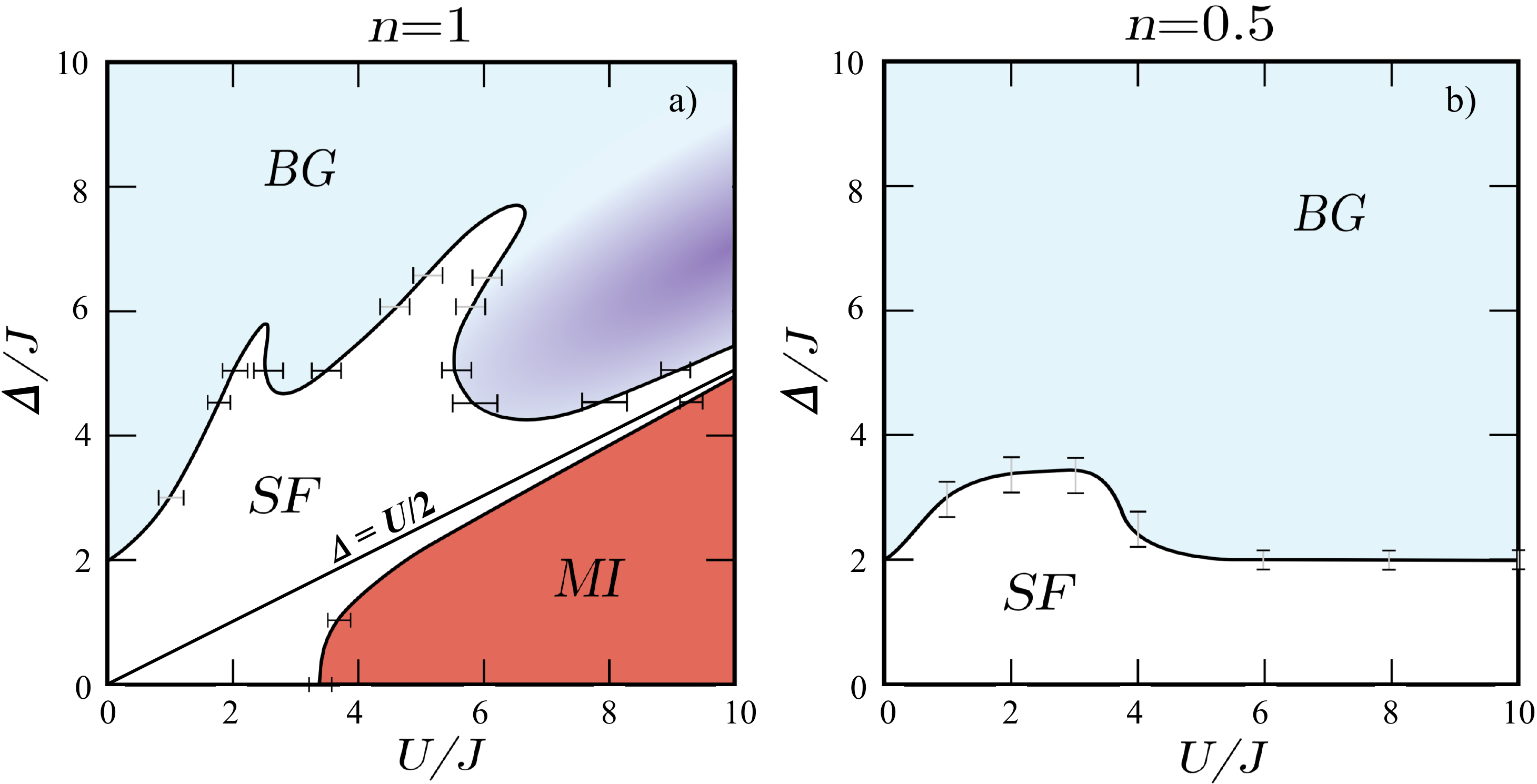}
    \caption{Phase diagrams for a quasiperiodic Bose--Hubbard model for densities $n=1$ (\textbf{a}) and $n=0.5$ (\textbf{b}). Figure adapted from Reference \cite{Roux2008}.} 
    \label{fig:theo1}
\end{figure}

The disorder potential $\epsilon_j$ can be generated in several ways, resulting in a specific spectral distribution. Both theoretically and experimentally, two cases are the most relevant: (a) random distribution of energies $\epsilon_j \in [-\Delta,\Delta]$ and (b) quasiperiodic distribution $\epsilon_j = \Delta \cos(2\pi j \sigma)$ with $\sigma$ being an irrational number~\cite{AubryAndre}. The latter can be experimentally generated by superimposing to a main periodic potential an auxiliary lattice one with incommensurate wavelength ($\lambda_2=\lambda_1/\sigma$). Hence, the three main energy scales characterizing the Hamiltonian, {i.e.,} the tunneling energy $J$, the quasidisorder strength $\Delta$, and the interaction energy $U$, can be controlled by tuning the depth of the main lattice $S_1$, the depth of the secondary one $S_2$ (being $\Delta = \sigma^2 S_2/2$), or the interparticle scattering length on a Feshbach resonance \cite{Chin2010}, respectively.
The simultaneous presence of disorder and commensurate potential generates a competition between the three possible quantum phases, namely the superfluid (SF) phase, the Mott insulator (MI), which occurs at large interactions for commensurate fillings, and the so-called Bose glass (BG) phase, which is induced by disorder. Figure~\ref{fig:theo1} shows the zero-temperature ($T = 0$) phase diagram of 1D bosons in the quasiperiodic lattice as a function of the ratios $\Delta/J$ and $U/J$, obtained by numerically solving the Bose--Hubbard problem~\cite{Roux2008}.

We must distinguish between two physically different situations depending on the average boson occupation number $n=N/M$, where  $N$ is the number of bosons and $M$ is the length of the 1D system. For incommensurate fillings ($n<1$), the system is similar to the continuum case~\cite{Ristivojevic2021}, with a SF phase replaced by a BG phase by increasing the disorder strength $\Delta$.  On the other hand, for unit filling ($n = 1$), the ground state is a MI with a gap. Adding disorder, such a gapped phase persists up to the value $\Delta = U/2$ (dashed line) where the excitation spectrum becomes gapless and the system first becomes a SF and then a BG. 

One-dimensional disordered bosonic systems, as described by Equation~\eqref{eq:BHdis}, provide an ideal platform for testing and developing precise theoretical methods for studying many-body physics, which yields useful predictions about the position of quantum phase transitions and the indications on the most appropriate observables for their detection. In the case of pseudorandom quasiperiodic disorder, phase diagrams have been obtained by exact numerical results on the Bose--Hubbard model in small systems~\cite{Roth2003,Bar-Gill2006}. However, they suffer a limited accuracy in locating the points of phase transition. More detailed results have been found by means of quantum Monte Carlo methods~\cite{Roscilde2008} and the density-matrix renormalization group (DMRG) algorithm~\cite{Roux2008}. These theoretical works represent the groundwork for the experimental detection of disordered quantum phases, and they also point out the experimental tools to detect the quantum phase transitions of these systems. In particular, beside the compressibility, Reference~\cite{Roux2008} points out to the measurement of \textit{coherence} of the quantum gas, which is detected by time-of-flight imaging from the width of the momentum distribution (see for results Section~\ref{sec:coh}). Another interesting tool for detecting phase transition is the observation of \textit{excitation spectrum} (see Section~\ref{sec:excitation}). The excitation spectrum of strongly repulsive 1D bosons in a disordered or quasiperiodic optical lattice has been computed~\cite{Orso2009}. The predicted excitation spectrum shows a peculiar behavior with two excitation peaks, one as expected around the repulsion energy scale $U$ with width $\sim 2\Delta$ and the other one centered at $\Delta$ with the same width. The prediction of the presence of an absorption feature in the low-frequency band appears as a consequence of the formation of a Bose glass at incommensurate filling, thus making the excitation spectrum measurement an important tool of investigation. Experimentally, it can be easily assessed by coherent lattice modulation spectroscopy~\cite{Fallani2007,Tarallo12}.


\section{Experimental Results}

\begin{figure}[t]
    \includegraphics[width=0.38\textwidth]{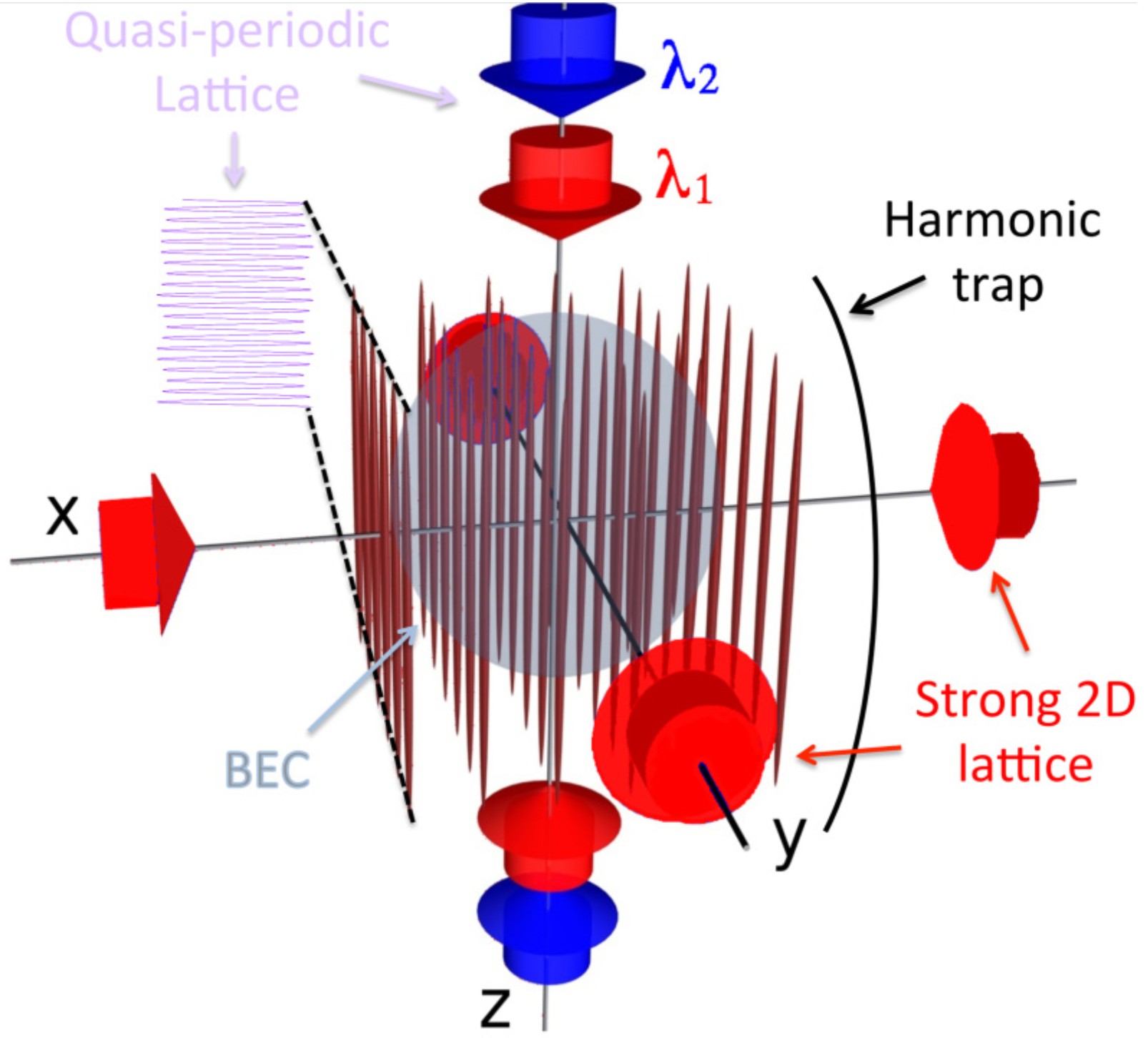}
    \caption{Schematic drawing of the typical experimental realization of a 1D disordered bosonic system. Two strong optical lattices are used to provide a tight confinement and form an array of 1D potential tubes. The axial quasiperiodic potential is formed by superimposing two incommensurate optical lattices of wavelengths $\lambda_1$ and $\lambda_2$. The harmonic trap results from the intensity gradient of the Gaussian laser beams.}
    \label{fig:Setup}
\end{figure}

The experimental realization of a 1D bosonic system with ultracold gases is schematically shown in Figure~\ref{fig:Setup}. Starting from a 3D Bose--Einstein condensate (BEC), the atoms are typically loaded in a strong 2D optical lattice~\cite{Bloch2005}. This traps the atoms to an array of tightly confining 1D potential tubes, thus generating a set of many quasi-1D systems. Along the 1D tubes, another optical lattice is employed to produce a set of disordered quasi-1D systems, which are described by the disordered Bose--Hubbard Hamiltonian in Equation~\eqref{eq:BHdis}. Here, the disorder is introduced either with a secondary optical lattice, generating the quasiperiodic disordered lattice~\cite{Fallani2007}, or by a second atomic species as system impurity~\cite{Gadway2011}. 

A systematic experimental study of the many-body properties of such a system can be performed by momentum distribution or by excitation energy measurements. The coherence (Section~\ref{sec:coh}) and transport (Section~\ref{sec:tra}) properties of the tubes can be studied by measuring the momentum distribution of the system, achieved through absorption imaging after a free expansion. These measurements correspond to an average over all the tubes of the systems, and thus over its different densities. In the case of transport measurements, to induce a dynamics on the atoms along the tubes, the system is brought out of equilibrium by a sudden change of the harmonic trap. The excitation spectra of the system are obtained by modulating the amplitude of the main lattice depth. The amount of energy absorbed by the system can be extracted by temperature measurements of the 3D BEC, which is recreated after an adiabatic switch-off of the 1D confinement \cite{DErrico2014}. Alternatively, the modulation heating effect can be detected by phase coherence measurements. Phase coherence is restored by reducing the depths of the trapping lattices to less than five recoil energies \cite{Gerbier2005}, while phase interference is imaged after a time-of-flight. Typically, the amount of heating can be quantified either by the visibility $\mathcal{V}$ of the interference peaks \cite{Gadway2011}, which is defined analogously to the optical case as a function of the atomic density $\rho$
$$
\mathcal{V} = \frac{\rho_{max}-\rho_{min}}{\rho_{max}+\rho_{min}},
$$
or by the width of the central peak \cite{Fallani2007,Guarrera2007} (see Section~\ref{sec:excitation}).


\subsection{Coherence}\label{sec:coh}

\begin{figure*}[tb]
\includegraphics[width=12.5 cm]{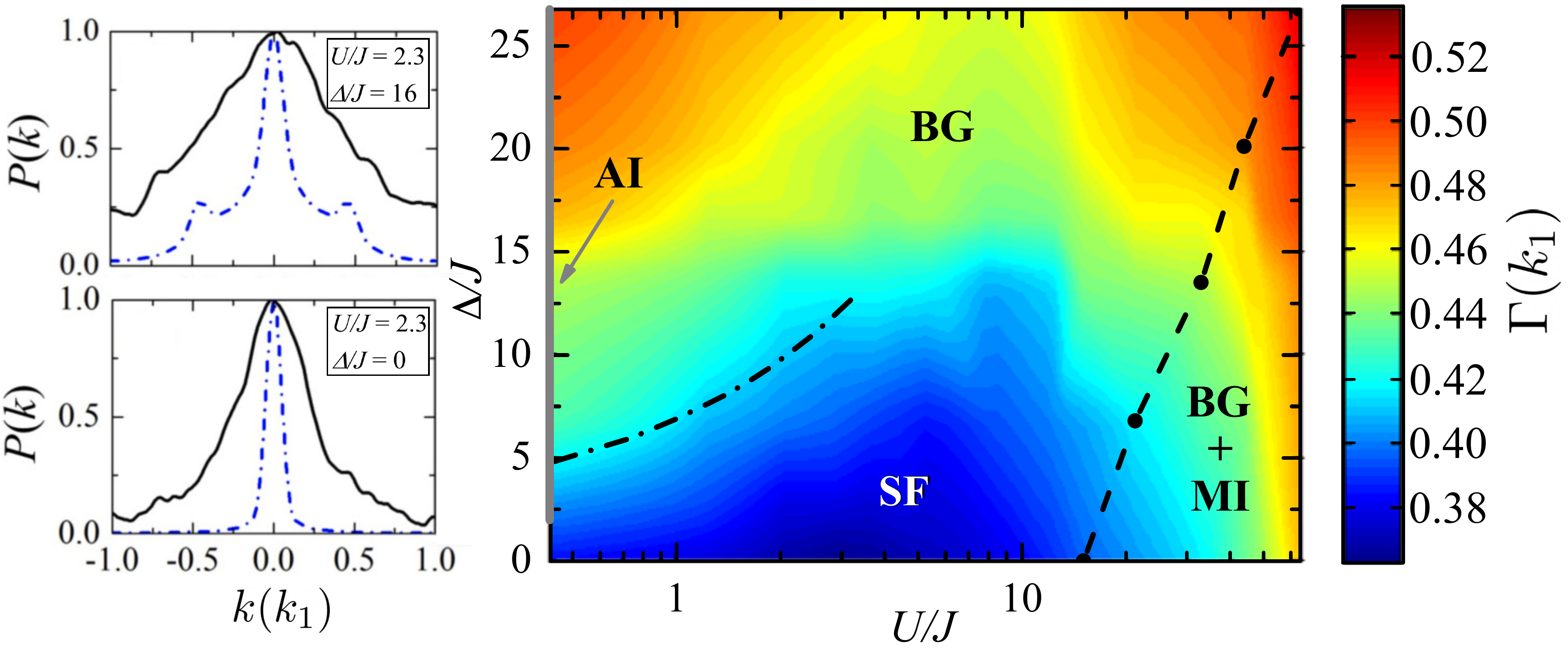}
\caption{\label{fig:Delta_U_Diagram} Measured rms width $\Gamma$ of the momentum distribution $P(k)$ of arrays of quasi-1D samples of $^{39}$K atoms in the $U-\Delta$ diagram. $\Gamma$ is reported in units of $k_1$, where $k_1=2 \pi / \lambda_1$ is the main lattice wavevector. The dashed line indicates the upper bound for the existence of the MI, according with $T=0$ DMRG calculations. Left panels show the measured momentum distribution (solid black line) for two points of the diagram in the SF (bottom) and BG (top) regions, compared with the $T=0$ DMRG calculations (dashed-dotted blue line). Figure adapted from Reference~\cite{DErrico2014}.}
\end{figure*}

An overview of the nature of a disordered interacting system has been provided by measurements of the momentum distribution $P(k)$ in an array of 1D tubes of $^{39}$K atoms in a quasiperiodic lattice. 
An experimental measurement of the coherence of the system is shown in Figure~\ref{fig:Delta_U_Diagram}, where the width $\Gamma$ of $P(k)$ is plotted as function of the interaction strength $U$ and the disorder strength $\Delta$.
At small $\Delta$ and $U$, the observation of a narrow $P(k)$ is a signature of a coherent regime (blue zone). For increasing values of the two energy scales, the coherent regime is progressively replaced by a more incoherent regime (green, yellow, and red zones). 
The observed increase of $\Gamma$ can be attributed to either the emergence of an insulating phase or to an increase in the temperature. The latter effect on $\Gamma$ has been experimentally excluded by entropy measurements~\cite{DErrico2014}. In fact, the measured entropy does not show any increase with increasing disorder strength. This suggests that the increased $\Gamma$ is due to the emergence of an insulating phase, as predicted for the $T=0$ temperature case.
Despite the finite $T$ and the inhomogeneity of the experimental tubes, the diagram behavior resembles that of the $T=0$ theoretical predictions for homogeneous systems, where the existence of a BG phase is predicted~\cite{Giamarchi&Schulz,Fisher&Fisher,Roscilde2008,Roux2008}.

Comparing the experimental diagram with the theory in Figure~\ref{fig:theo1}, we see that for increasing interaction along the $\Delta=0$ line,  $\Gamma$ increases due to the progressive formation of an incoherent MI. For increasing disorder along the $U=0$ line, the system forms an Anderson insulator for $\Delta>2J$ \cite{Modugno_2009}. For weak disorder and interaction, the system is in a SF regime, surrounded by a re-entrant insulating regime extending from small to large $U$. In the weakly interacting regime, a crossover from the incoherent disorder-induced insulator toward more coherent regimes is observed when the interaction energy $n U \gtrsim \Delta-2J$ (see dashed-dotted line in Figure~\ref{fig:Delta_U_Diagram}). In the strongly interacting regime, disorder and interactions cooperate to localize the system and a second crossover towards less coherent regimes occurs. The interaction induced MI, which for a homogeneous system with $n=1$ is expected to survive in the disordered potentials only for moderate disorder $\Delta < U/2$, is expected to exist in the experimental inhomogeneous one only below the dashed black line shown in Figure~\ref{fig:Delta_U_Diagram}. In this region, as an effect of the inhomogeneous density of the experimental system, for $\Delta<2J$, the MI coexists with a SF fraction, which is localized by the disorder in a BG phase for $\Delta>2J$. 

For a complete comparison of the experimental phase diagram with theoretical predictions, it would be necessary to include both finite temperature and inhomogeneity of the experimental system into numerical simulations. This would result in costly numerical calculations. If only system inhomogeneity is included, zero-temperature DMRG calculations (left panels in Figure \ref{fig:Delta_U_Diagram}) find a diagram with a general behavior close to the experimental one but with a SF $\Gamma$ much smaller than that observed in the experiments~\cite{DErrico2014}. To include the finite temperature of the experimental system, two different DMRG schemes have been developed: (i) a direct simulation of the thermal density matrix in the form of a matrix-product purification and (ii) a less costly phenomenological method based on DMRG ground-state data that are extended to finite temperatures by introducing an effective thermal correlation length~\cite{Gori2016}. These simulations have shown that, while in the weakly interacting regime thermal effects can be rather strong, they are significantly less relevant in the strongly interacting one. There, the scaling of the correlation length with $T$ shows a weak dependence below a crossover temperature, indicating that the strongly correlated quantum phases predicted by the $T = 0$ theory can persist at finite temperatures.

\subsection{Transport}\label{sec:tra}

\begin{figure*}[b]	
\includegraphics[width= 12 cm]{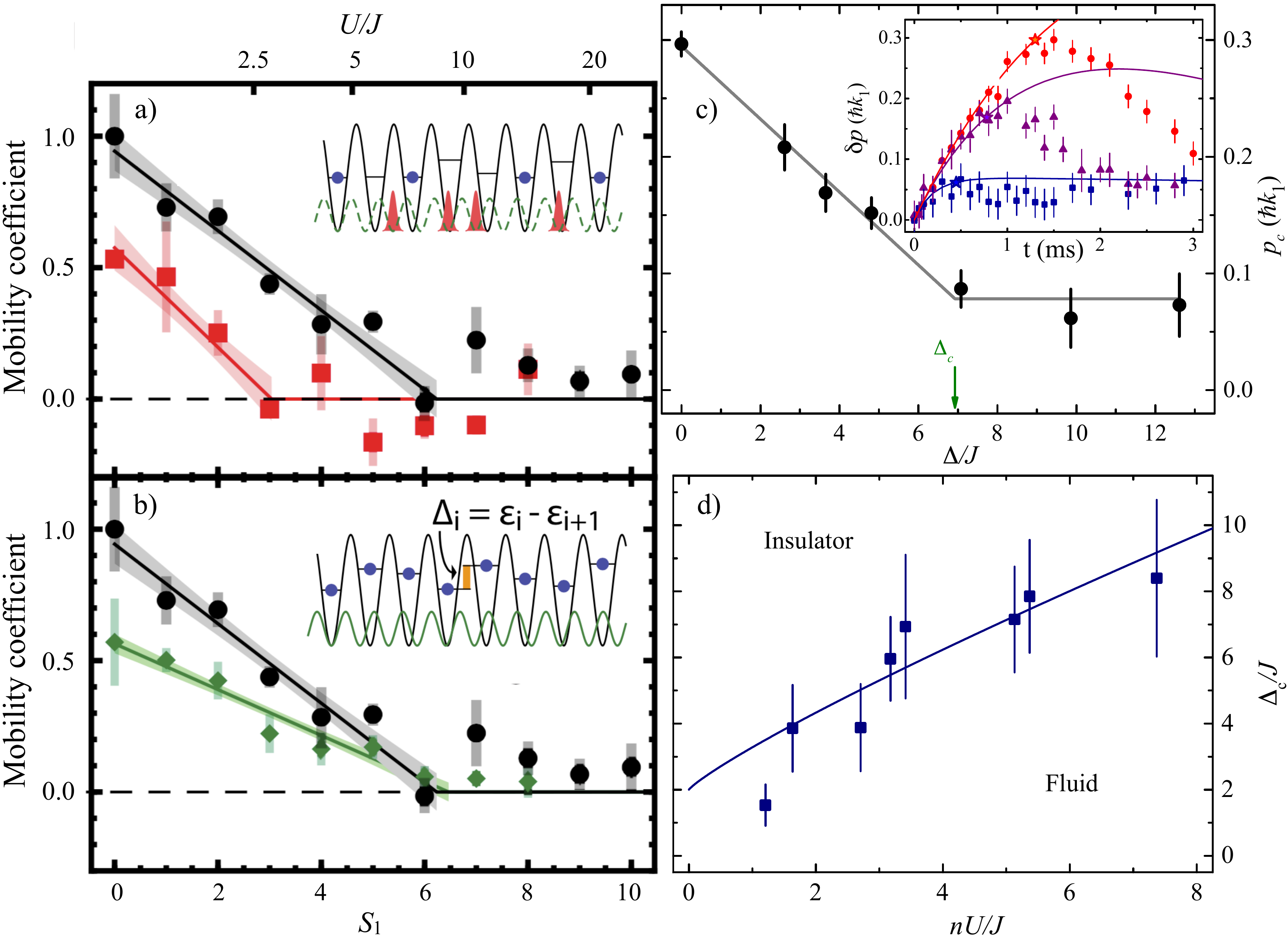}
\caption{\label{fig:SF_Ins_Transition} (\textbf{a},\textbf{b}) {Mobility coefficient} of $^{87}$Rb after impulse excitation versus lattice depth $S_1$ without disorder (black circles) compared with different disorder configurations: (\textbf{a}) atomic impurities ($f_{imp}= 0.5$, red squares) and (\textbf{b}) an incommensurate lattice ($S_2=3$, green diamonds). The lattice depths are defined in units of recoil energies, $S_i=V_i/E_{ri}$ with $i=1,2$. Figure adapted from Reference~\cite{Gadway2011}. (\textbf{c},\textbf{d}) Transport after trap excitation of an array of quasi-1D samples of weakly interacting $^{39}$K atoms: (\textbf{c}) critical momentum $p_c$ for weak interaction ($U/J=1.26$) as a function of the disorder strength and (\textbf{d}) critical $\Delta_c/J$ at the fluid--insulator transition in the disorder-interaction plane, extracted by several piecewise fits of $p_c$ as a function of $\Delta$ for different values of fixed $U$ (solid line in panel (\textbf{c})). In the inset of panel \textbf{c}, the time evolution of $\delta p$, from which $p_c$ has been extracted (stars), is shown for three values of $\Delta$: $\Delta = 0$ (red circles), $\Delta = 3.6 J$ (purple triangles), and $\Delta=10 J$ (blue squared). Figure adapted from Reference \cite{Tanzi2013}.}
\end{figure*}

The insulating nature of the incoherent region has been confirmed by transport measurements. The mobility can be measured by observing the system evolution after an impulse has been applied to it. In Figure~\ref{fig:SF_Ins_Transition}a,b, the results from the first experiments with $^{87}$Rb atoms are shown. The clean case ($\Delta=0$) is compared with two different disordered configurations: atomic impurities (Figure~\ref{fig:SF_Ins_Transition}a) and quasiperiodic potential (Figure~\ref{fig:SF_Ins_Transition}b). When a variable impulse is applied to the system, the velocity acquired by the atoms can be fitted with a linear function whose slope defines the mobility coefficient. In the absence of disorder, the mobility coefficient decreases with the increasing in the potential lattice depth $S_1$ and reaches zero mobility when entering in the MI region. When disorder is present, the behavior is analogous, suggesting the system is entering in an insulating regime. Nevertheless, while with impurities the transition to the zero-mobility was shifted towards smaller values of $S_1$, in the case of the quasiperiodic potential (with constant $S_2$), no shift of the critical depth is measured. Such different behavior could be due to the fact that increasing $S_1$ towards the insulating regime, the disorder $\Delta/J$ is decreasing, thus pushing the critical interaction to enter the BG regime to larger values of $U/J$, where the BG phase coexists with the MI one. This problem has been bypassed in $^{39}$K experiments by using Feshbach resonances to tune the interaction independently from the value of $S_1$~\cite{Roati2007,DErrico2007}. 

Figure~\ref{fig:SF_Ins_Transition}c,d shows momentum dependent transport measurements in the weakly interacting regime. The experimental protocol consists of tracking the time evolution of the momentum $\delta p$ acquired by the system for different values of the disorder strength $\Delta$ and the interaction energy $U$, tuned via Feshbach resonance. Typical datasets of such measurements are plotted in the inset of Figure~\ref{fig:SF_Ins_Transition}c, where $\Delta$ is different for each dataset, while $U$ is kept constant. Here, we can observe that the system explores a sharp transition from a weakly dissipative regime (at small $\delta p$), well fitted with a damped oscillation function (solid lines), to a strongly unstable one (at large $\delta p$). The critical momentum $p_c$ separating the two regimes has been identified as the momentum value, where the experimental data deviate from the fitting curve used in the first regime (stars in the inset of Figure~\ref{fig:SF_Ins_Transition}c). The measured critical momentum $p_c$ at each $\Delta$, similarly to the previously described mobility coefficient, linearly decreases until it reaches a plateau value, corresponding to the insulating regime of the system. With a piecewise fit of $p_c$, one can extract the critical disorder strength $\Delta_c$ to enter in the insulating regime at fixed interaction energy $U$ (Figure~\ref{fig:SF_Ins_Transition}c). Repeating the measurements for different interactions, it has been observed that the critical disorder to enter the insulating regime increases with $U/J$ (Figure~\ref{fig:SF_Ins_Transition}d), at least for weak interaction. By employing the vanishing of $p_c$ for the observed instability the fluid--insulator transition driven by disorder has been located, across the interaction-disorder plane in the weakly interacting regime. In fact, while the experiments with $^{87}$Rb atoms are limited to the strongly interacting regime, the momentum-dependent measurements with $^{39}$K samples allow researchers to investigate the weakly interacting one.
 
\begin{figure*}[tb]
\includegraphics[width= 13 cm]{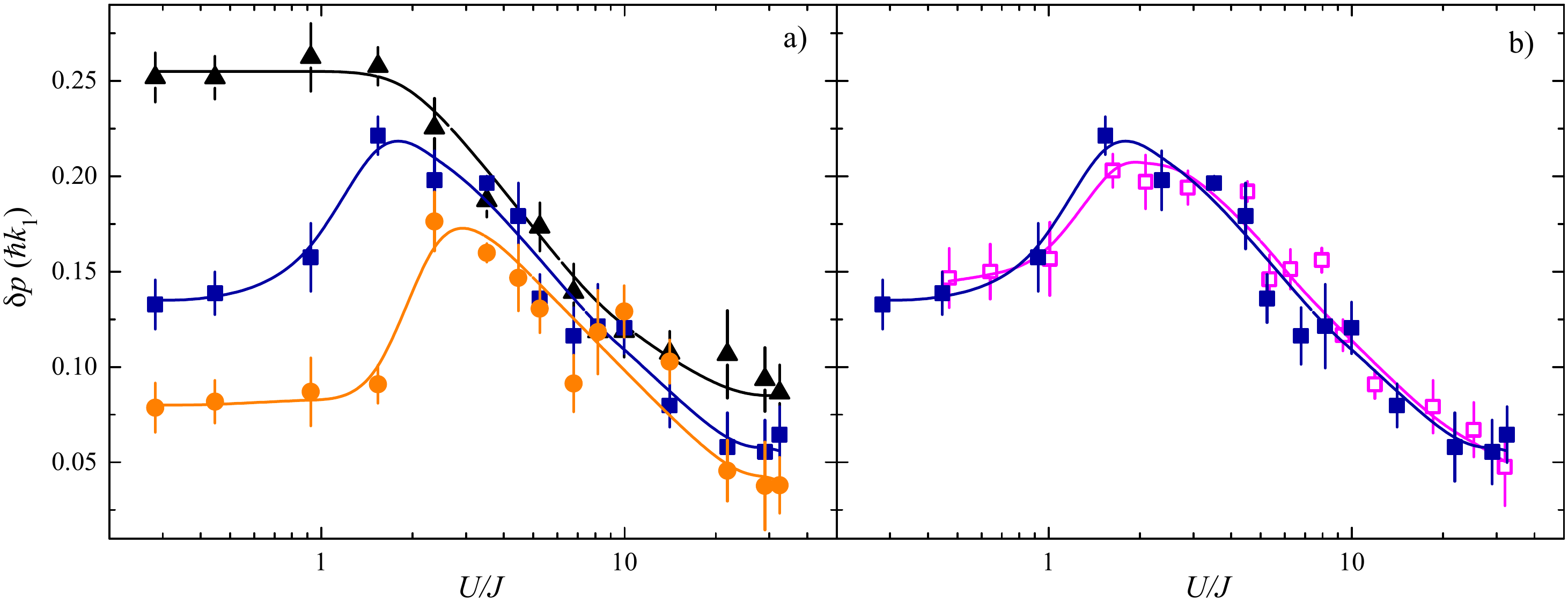}
\caption{\label{fig:Mobility_K} {Effective mobility} as a function of disorder and interactions. (\textbf{a}) Momentum $\delta p$ acquired by the system after a fixed evolving time $t = 0.9$ ms in the tilted potential for three different disorder strengths, $\Delta = 0$ (black triangles), $\Delta = 6.2J$  (blue squares), and $\Delta = 8.8J$ (orange circles). (\textbf{b}) The $\Delta = 6.2J$ measurements are acquired for two temperatures of the SF component, $k_B T = 3.1(4)J$ (full blue) and $k_B T=4.5(7)J$ (empty magenta).  Figure adapted from Reference \cite{DErrico2014}.}
\end{figure*}

In order to confirm the insulating nature of the observed incoherent regimes in the full diagram of Figure~\ref{fig:Delta_U_Diagram}, the momentum $\delta p$ acquired by the $^{39}$K system after a fixed time from its excitation has been measured (Figure~\ref{fig:Mobility_K}). 
This effective mobility is shown in \mbox{Figure~\ref{fig:Mobility_K}a} for the clean case and for two fixed values of the disorder strength. With no disorder and small $U$ the system is conductive, while the mobility decreases when approaching the MI region. With finite disorder, instead, the system is insulating for both very weak and strong interactions, while a finite mobility can be recovered for moderate values of $U$. These results indicate that the incoherent regimes at both weak and strong $U$ are also insulating, thus confirming the re-entrant behavior of the insulating regime observed in the coherence diagram. An additional measurement performed at a higher temperature indicates that, as expected by theory \cite{Aleiner2010}, the mobility for intermediate disorder strength is essentially $T$-independent in the explored range $k_B T = (3.1$--$4.5)J$ (Figure~\ref{fig:Mobility_K}b).

\subsection{Excitation Spectra}\label{sec:excitation}

\begin{figure*}[tb]
\includegraphics[width= 14 cm]{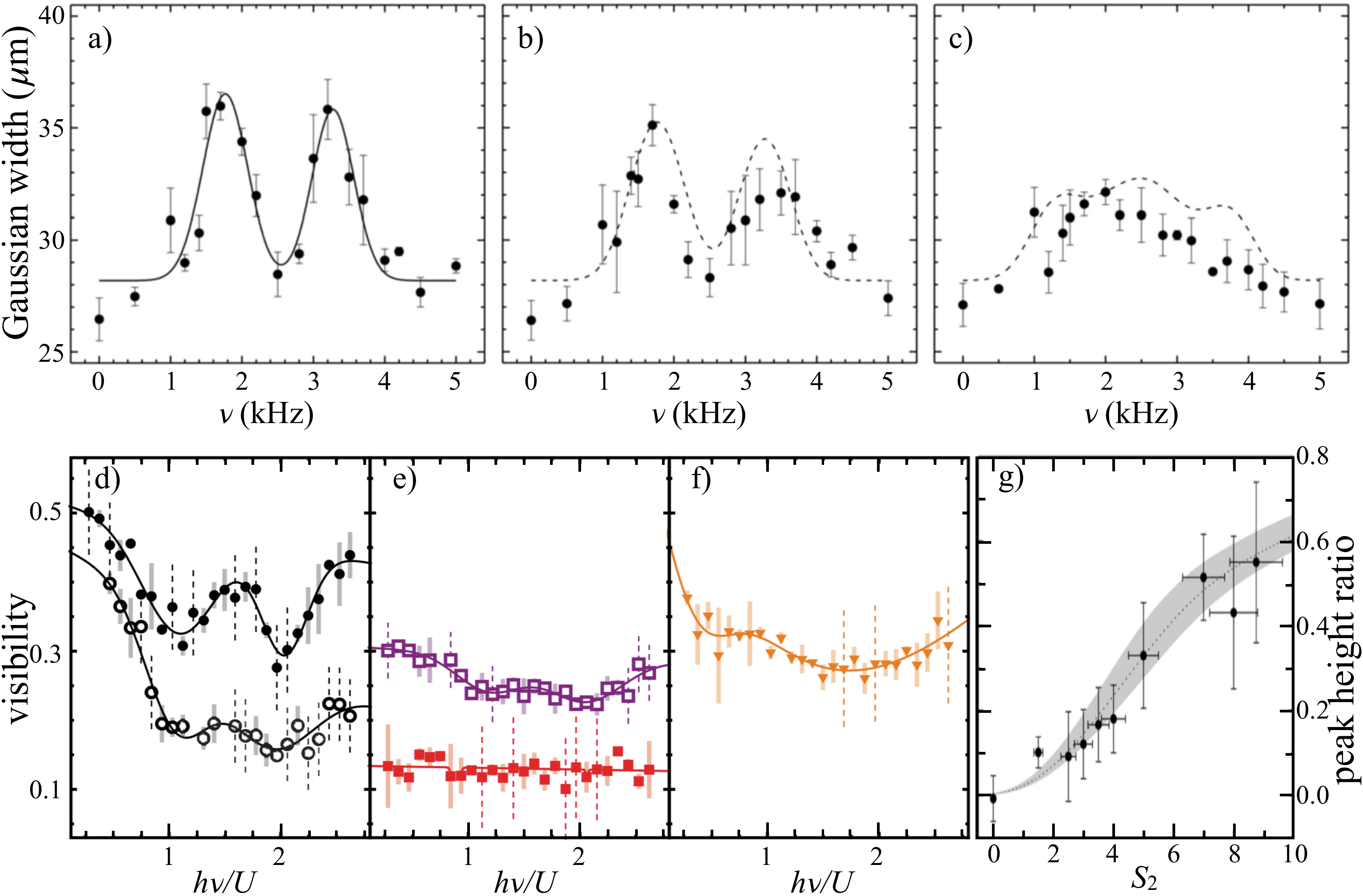}
\caption{\label{fig:Spettri_Rb} {Excitation spectra} (\textbf{a}--\textbf{f})~and noise correlations spectroscopy (\textbf{g}) of arrays of quasi-1D samples of $^{87}$Rb atoms. (\textbf{a}--\textbf{c})~Spectra in a quasiperiodic potential for a depth $S_1=25$ of the main lattice and increasing depths of the secondary lattice: (\textbf{a}) $S_2=0$ , (\textbf{b}) $S_2=0.2$ , and (\textbf{c}) $S_2=0.5$. Here, the spectra have been quantified by measuring the Gaussian width of the central peak of the momentum distribution of atoms released from the lattices with reduced intensity $(S_1=5$, $S_2=0$). Figure adapted from Reference \cite{Guarrera2007}. (\textbf{d}--\textbf{f}) Visibility of excitation gap for different disorder configurations: (\textbf{d}) in the absence of disorder for $S_1=9$ and $S_1=14$ (open and filled black circles), (\textbf{e}) for $S_1=14$ with atomic impurity fractions $f_{imp}=0.1$ (open purple squares) and  $f_{imp}=0.5$ (filled red squares), and (\textbf{f}) for $S_1=14$,  with no impurities and an incommensurate lattice of depth $S_2=1$ (orange triangles). Here, the spectra have been quantified by measuring the interference peak visibility in the momentum distribution of atoms released from the lattice with reduced intensity $(S_1=4$, $S_2=0$) \cite{Gadway2010}. Figure adapted from Reference \cite{Gadway2011}. (\textbf{g}) Noise correlation spectroscopy in a quasiperiodic potential: the ratio between the height of the $k_2$ and $k_1$ correlation peaks as a function of $S_2$. Figure adapted from Reference \cite{Guarrera2008}. The lattice depths are defined in units of recoil energies, $S_i=V_i/E_{ri}$ with $i=1,2$.}
\end{figure*}

To probe the nature of the insulating phases, it is necessary to investigate the excitation properties of the system. This can be undertaken by performing lattice modulation spectroscopy, {i.e.}, by measuring the energy absorbed by the system after a sinusoidal amplitude modulation of the main lattice at fixed frequency $\nu$. While the MI is known to be gapped, the BG phase is predicted to be a gapless insulator. 
First experiments with $^{87}$Rb observed the broadening of the typical MI spectrum (Figure~\ref{fig:Spettri_Rb}), with both quasiperiodic potentials \cite{Fallani2007, Gadway2011} and localized impurity atoms \cite{Gadway2011}. Despite showing signatures of BG formation, they do not permit to distinguish a specific signature of the BG spectrum due to the strong interaction ($U>50J$) and the strong disorder ($\Delta>50J$) regime. Moreover, noise correlation spectroscopy allowed experimentalists to monitor the destruction of the MI ordered structure in the presence of an additional secondary lattice potential (Figure~\ref{fig:Spettri_Rb}g), but not to highlight a specific feature due to the BG phase.


\begin{figure*}[tb]
\includegraphics[width= 14 cm]{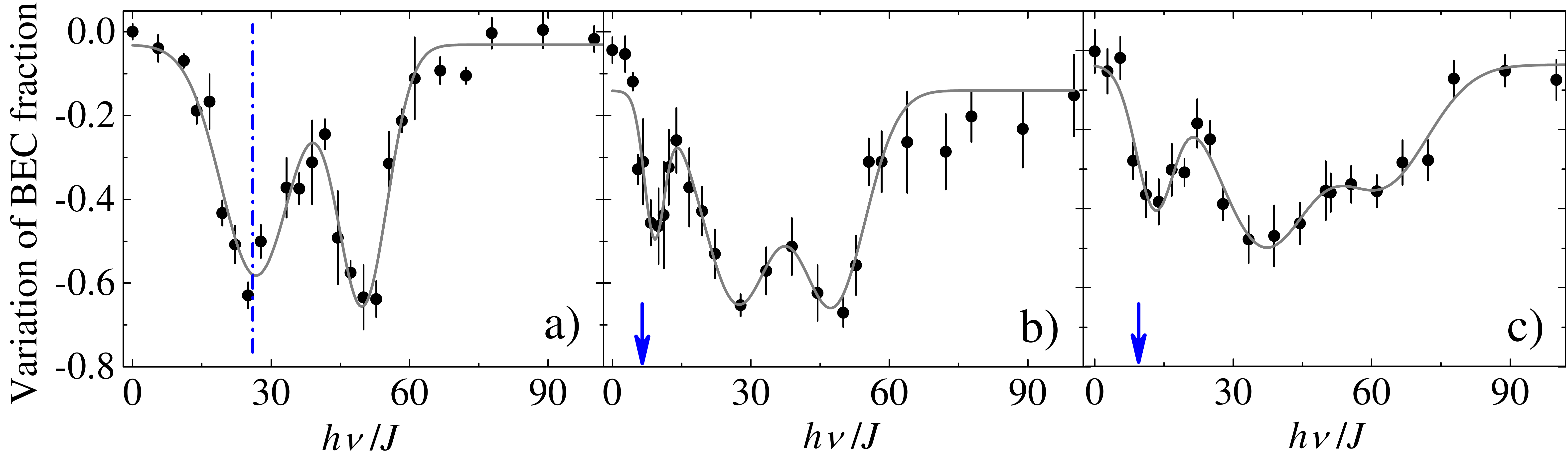}
\caption{\label{fig:Spettri_K} {Excitation spectra} of arrays of quasi-1D samples of $^{39}$K atoms with strong interactions. (\textbf{a}--\textbf{c}) Experimental spectra for $U=26J$ and $\Delta=0$ (\textbf{a}), $\Delta=6.5J$ (\textbf{b}), and $\Delta=9.5J$ (\textbf{c}). The spectra have been quantified by measuring the relative variation of the BEC fraction with respect to the unexcited value ($\nu=0$). The blue arrows are at $h \nu = \Delta$, the dashed-dotted line in (\textbf{a}) is at $h \nu =U$, and the continuous lines are fits with multiple Gaussians. 
Figure adapted from Reference \cite{DErrico2014}.}
\end{figure*}

Experiments with $^{39}$K permit to explore the excitation spectrum in the full range of interaction and disorder diagram and to find regions where it is possible to distinguish the gapless spectrum of the BG from the gapped one of the MI. In these experiments, the absorbed energy has been quantified by measuring the temperature of the BEC after the adiabatic switch-off of the 1D confinement. Depending on the amount of acquired energy, the time-of-flight atomic distribution can be fitted either by a two-component function (a Thomas--Fermi profile plus a Gaussian distribution) or by a Gaussian function. In the former case, the heating is related to the BEC fraction; in the latter, it is related to the width $\sigma$ of the Gaussian distribution. Let us start from the strongly interacting regime, where, in the presence of moderate disorder, the BG phase should coexist with the MI (Figure~\ref{fig:Spettri_K}). In the clean case, the spectrum is characterized by the double peak shape typical of the trapped MI, with a first peak centered at $h \nu =U$ due to excitation between sites in the MI domains with the same filling. In addition, a second peak is centered at $h \nu = 2U$ due to excitation between sites in the MI domains with different occupations. Adding a finite disorder, the spectrum shows a clear change. First, there is a broadening of the MI peaks, as already observed with $^{87}$Rb experiments at strong disorder. Second, at low frequencies, it appears an extra peak filling the Mott gap, centered around $h \nu =\Delta$, which can be ascribed to the regions with incommensurate filling, {i.e.,} to the BG phase. 


\begin{figure*}[tb]
\includegraphics[width= 12.5 cm]{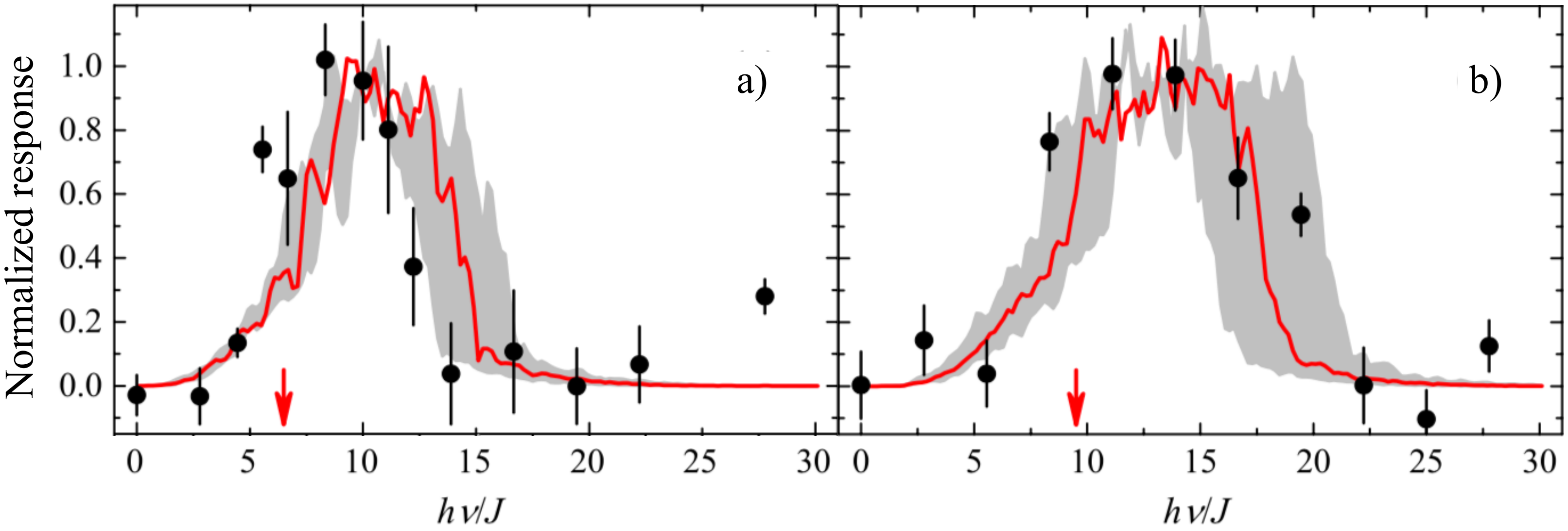}
\caption{\label{fig:Spettri_BG} {Excitation spectrum} of the strongly interacting BG. The experimental data (black circles) for the low-frequency part of the spectra in Figure~\ref{fig:Spettri_K} are compared with theory (red solid line) for two disorder strengths: (\textbf{a}) $\Delta=6.5J$ and (\textbf{b}) $\Delta=9.5J$. The grey region shows the effect of a 20\% uncertainty on $\Delta$. The red arrows are at $h \nu = \Delta$. Figure adapted from {Supplemental Materials} of \mbox{Reference~\cite{DErrico2014}}.}
\end{figure*}

The agreement between BG theory and experiment is best understood once the MI background is subtracted from the experimental data. Figure~\ref{fig:Spettri_BG} shows a zoom of the excitation spectra around the disorder strength energy $\Delta$ after the Gaussian background of the MI peak has been subtracted, and the resulting peak response has been normalized to unity. We can see that the experimental spectra of the BG are reasonably well reproduced by theory calculations, where a fermionized-boson model has been used~\cite{Orso2009,Pupillo2006}. 

\begin{figure*}[tb]
\includegraphics[width= 15 cm]{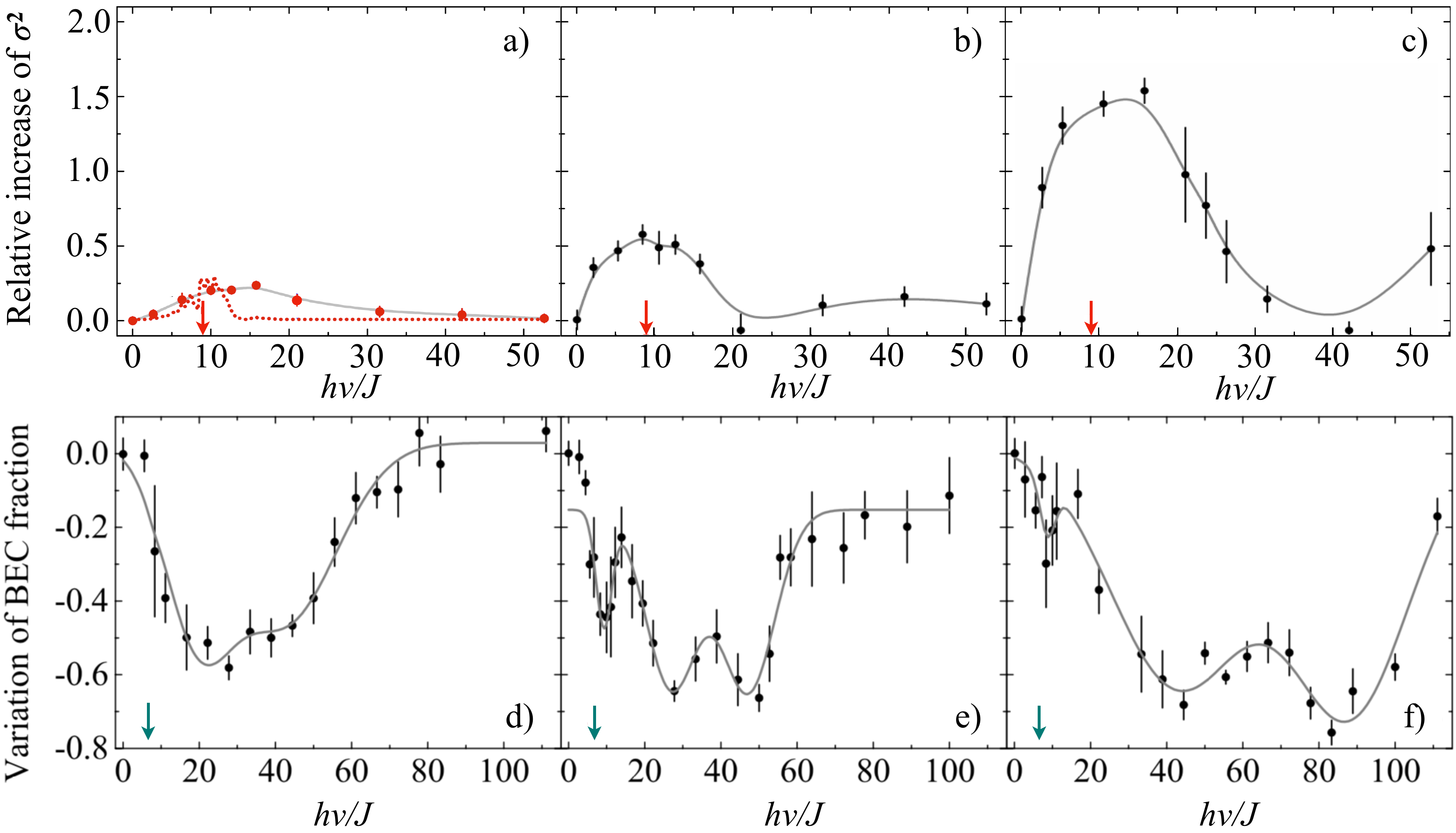}
\caption{\label{fig:SpettroK_vs_U} {Excitation spectra} of $^{39}$K atoms from weak to strong interactions. (\textbf{a}--\textbf{c}) Excitation spectra for fixed disorder, $\Delta=8.9J$, and small increasing interactions: $U=0.35J$ (\textbf{a}), $U=1.4J$ (\textbf{b}), or $U=2.1J$ (\textbf{c}). The spectra have been quantified by measuring the relative increase of $\sigma$ with respect to the unexcited value ($\nu=0$). (\textbf{d}--\textbf{f}) Excitation spectra for fixed disorder, $\Delta=6.5J$, and large increasing interactions: $U=20J$ (\textbf{d}), $U=26J$ (\textbf{e}), or $U=58J$ (\textbf{f}). The arrows mark $\Delta/J$. The spectra have been quantified by measuring the relative variation of the BEC fraction with respect to the unexcited value ($\nu=0$). Figure adapted from Reference \cite{DErrico2014}.}
\end{figure*}

We now analyze the spectral properties of the system across the phase diagram. In Figure~\ref{fig:SpettroK_vs_U}, the behavior of the excitation spectrum moving from weak to strong interaction at a given finite disorder is shown. In the case of weak interaction, the excitation spectra at $\Delta = 8.9 J$  are shown for three increasing values of $U$ (Figure~\ref{fig:SpettroK_vs_U}a--c). For vanishing $U$, a weak excitation peak centered at $\Delta$ has been observed, consistent with the presence of an Anderson insulator. The experimental excitation spectrum is well reproduced by a non-interacting bosonic model (Figure~\ref{fig:SpettroK_vs_U}a). Increasing $U$, the system response progressively enhances and broadens (Figure~\ref{fig:SpettroK_vs_U}b), ending up with an excitation spectrum that is undistinguishable from that of a clean SF (Figure~\ref{fig:SpettroK_vs_U}c). This behavior is thus consistent with the system crossing the BG--SF transition.

In the case of strong interaction, the excitation spectra at the $\Delta = 6.5 J$  are shown for three increasing values of $U$ (Figure~\ref{fig:SpettroK_vs_U}d--f). The peak centered at $\Delta$ is the signature of the strongly correlated BG. Such “$\Delta$-peak” can be observed only in a limited region of $\Delta$ and $U$ values. When $U$ is comparable with $\Delta$, the MI and BG peaks overlap, the former being typically larger and covering the latter (Figure~\ref{fig:SpettroK_vs_U}d). When $U$ is much larger than $\Delta$, the fraction of sites with incommensurate density that can form a BG becomes negligible and, again, only the MI peaks are clearly detectable (Figure~\ref{fig:SpettroK_vs_U}f). Furthermore, for very large disorder strengths ($\Delta > 20 J$), the spectrum becomes very broad and is only weakly affected by interaction, indicating that the system behavior is dominated by disorder, and any feature is observable.

The measurements of the excitation spectra, together with those of coherence (\mbox{Figure~\ref{fig:Delta_U_Diagram}}) and transport (Figure~\ref{fig:Mobility_K}), confirm an opposite
nature of the two regimes of weak and strong $U$, respectively, bosonic and fermionic, and an opposite role of the interactions. In the low-$U$ bosonic case, small repulsive interactions compete with disorder and screen the disorder-induced localization, favoring the coupling of single-particle states and gradually restoring coherence between particles and superfluidity.
In the large-$U$ fermionic case, instead, strong interactions induce fermonization of the bosonic sample, thus favoring, in the presence of disorder, Anderson localization.

\section{Outlook and Perspectives}
In this brief review, we discuss the experiments with 1D bosons where the effect of disorder has been investigated in the disorder-interaction plane. The topic of quantum matter in the presence of disorder is very complex, in particular, when dealing with experimental systems being inhomogeneous and at finite temperature. The coexistence of fractions with different densities, in fact, transforms the theoretical sharp quantum phase transitions into broad crossovers. A way to overcome this limit in future experiments could be to use a flat-top beam shaper providing homogeneous trapped systems \cite{Veldkamp82,Hoffnagle00,Tarallo07,Liang09,Gaunt2013}. This should also allow, in the strongly interacting regime, for a better discrimination of the BG and the MI phases. Concerning the problem of the finite temperature, it would be very important to reduce the actual temperature of the atomic 1D systems. The main source of heating is typically the phase noise affecting the 2D strong radial lattices and the main axial one. Recent theoretical calculations suggest to use a shallow quasiperiodic potential to reduce the lattice heating effect without losing information about the underlying quantum phases~\cite{Yao2020}. Another possibility could be to apply a phase stabilization on the lattices~\cite{Li2021}.

An intriguing direction of investigation would be the direct study of the effect of temperature on 1D disordered phases. A possible experimental implementation consists of using a second BEC insensitive to the lattices as a thermal bath \cite{Catani2009,McKay2013}. This would ensure both the thermal equilibrium in the 1D system and to have an independent measure of its temperature.

Another interesting question related to disordered systems is whether the existence of the finite temperature insulating phase in the weakly interacting regime could be related to the hot topic of many-body localization \cite{MBL_Review2019}. Different experiments with ultracold atoms recently investigated the many-body localization phenomenon, mainly for fermions \cite{MBL_Bloch2015,MBL_Bloch2016,MBL_Monroe2016,MBL_Bloch2017}, and only later for a disordered Bose--Hubbard system \cite{MBL_Greiner_Nature,MBL_Greiner_Science,MBL_Greiner_ArXiv}, but its existence is still under debate \cite{Panda_2020}.



\vspace{6pt}

\bibliography{biblio.bib}
\end{document}